\documentclass[showpacs,preprintnumbers,amsmath,amssymb]{revtex4}

\def\ben{\begin{enumerate}} \def\een{\end{enumerate}}
\def\beq{\begin{equation}} \def\eeq{\end{equation}}
\def\bea{\begin{eqnarray}} \def\eea{\end{eqnarray}}
\def\beann{\begin{eqnarray*}} \def\eeann{\end{eqnarray*}}
\def\beasn{\begin{sneqnarray}} \def\eeasn{\end{sneqnarray}}
\begin{document}
%%%%%%%%%%%%%%%%%%%%%%%%%%%%%%%%%%%%%%%%%%%%%%%%%%%%%%%%%%%%%%%%

\date{2 March 2005}

%%%%%%%%%%%%%%%%%%%%%%%%%%%%%%%%%%%%%%%%%%%%%%%%%%%%%%%%%%%%%%%%
\title{The issue of time in generally covariant theories and the
Komar-Bergmann approach to observables in general relativity} %\author{}
 %\markright{}
%%%%%%%%%%%%%%%%%%%%%%%%%%%%%%%%%%%%%%%%%%%%%%%%%%%%%%%%%%%%%%%%

\author{J. M. Pons}

\affiliation{Departament d'Estructura i Constituents de la Mat\`eria\\
   Universitat de Barcelona,
and Institut de F\'\i sica d'Altes Energies,\\
   Av.~Diagonal 647,
   08028 Barcelona,
   Catalonia, Spain\\
   E-mail: pons@ecm.ub.es}

\author{D. C. Salisbury}

\affiliation{Department of Physics,
Austin College, Sherman, Texas 75090-4440, USA\\E-mail:
dsalisbury@austincollege.edu}
%%%%%%%%%%%%%%%%%%%%%%%%%%%%%%%%%%%%%%%%%%%%%%%%%%%%%%%%%%%%%%%%

%%%%%%%%%%%%%%%%%%%%%%%%%%%%%%%%%%%%%%%%%%%%%%%%%%%%%%%%%%%%%%%%
\begin{abstract}
Diffeomorphism-induced symmetry transformations and time evolution
are distinct operations in generally covariant theories formulated in phase space. Time is
not frozen. Diffeomorphism invariants are consequently not
necessarily constants of the motion. Time-dependent invariants arise through the choice of an intrinsic time, or equivalently through the
imposition of time-dependent gauge fixation conditions. One
example of such a time-dependent gauge fixing  is
the Komar-Bergmann use of Weyl curvature scalars in general
relativity.
An analogous gauge fixing is also imposed for the
relativistic free particle and the resulting complete set time-dependent
invariants for this exactly solvable model are displayed. In contrast with the free particle
case, we show that gauge invariants that are simultaneously
constants of motion cannot exist in general relativity. They vary with intrinsic time.
\end{abstract}

\pacs{4.20.Fy, 4.60.Ds. \hfill gr-qc/0503013}

%%%%%%%%%%%%%%%%%%%%%%%%%%%%%%%%%%%%%%%%%%%%%%%%%%%%%%%%%%%%%%%%
\maketitle
%\clearpage

\section{Introduction }

Generally covariant theories in phase space have in common that
the Hamiltonian is a linear combination of first class
constraints. This means that the Hamiltonian vanishes ``on
shell'', i.e., when the equations of motion are satisfied.
\footnote{This is no longer strictly true when boundary terms are
present as is the case in asymptotically flat spacetimes in
general relativity; but these terms do not affect time evolution
and will not be relevant to our discussion.}. Certain
combinations of first class constraints generate gauge
symmetries. And since rigid translation in time coordinate is a spacetime
diffeomorphism which does engender corresponding gauge symmetries
of dynamical variables in configuration-velocity space, some
authors have concluded that the Hamiltonian is itself a symmetry
generator. This interpretation has led to the claim that since
time evolution is just a gauge symmetry transformation there is
no real physical evolution of states in the classical canonical
formulation of generally covariant theories. \footnote{In this paper we consider only the classical "problem of time". Quantum aspects, such as the relation of time to the Wheeler-deWitt equation, will be dealt with in future publications.} So it would appear
that the canonical phase space approach encounters a disturbing
conceptual problem: if there is no physical time evolution a) the
theory seems to no longer coincide with the formulation in
configuration-velocity space and b) the very concept of time as
an evolutionary parameter seems to lose any meaning. This
assertion, that time evolution equals gauge symmetry, can be viewed
from other perspectives. For instance, it is encountered again
when one applies a gauge fixing (GF) and finds that the final
evolution generator vanishes; one then speaks of the frozen time
problem. Finally, a third view of the problem comes from the
definition of observables since the claim that time evolution is
gauge leads to the statement that the only possible observables
are constants of motion. Of course, this unsettling state of
affairs deserves careful scrutiny.

In this paper we
will show that there is no conceptual problem whatsoever for the
canonical formulation of generally covariant theories because the
mathematical identification of the Hamiltonian as a gauge generator is
erroneous. Briefly, the
Hamiltonian evolves solutions from their initial data;
the gauge generator,
as a symmetry of the equations of motion,
maps entire solution trajectories into new solution trajectories.
\footnote{Note that in the case of diffeomorphisms as gauge
transformations we will use thoroughout the active view for which
the spacetime coordinates are preserved whereas the
objects (metric and other fields) defined on them are transformed under
the diffeomorphisms.}

The distinction between time evolution and gauge symmetry can be
made in configuration-velocity space. But it is perhaps most
interesting in phase space since this is the arena in which one
hopes to make the canonical transition to quantum theory. We will
apply our remarks frequently to phase space; this perspective is
made possible by recent work in which it was shown how the
four-dimensional diffeomorphism induced gauge symmetry is
realized as a canonical transformation group on the full set of
canonical variables, including the lapse and shift
\cite{pss:1997pr,pss:2000jmp,pss:2000grg,pss:2000pr}.
Consequently we can demonstrate in detail the diffeomorpism
invariance of the phase space functions proposed originally by
Komar and Bergmann.

The point of view we advocate in this paper is consistent with
statements made by Don Marolf \cite{marolf95} and Carlo Rovelli
\cite{rovelli01a} regarding the nature
of diffeomorphism invariants. One of the merits of the present work is
that we describe precisely in what sense observables are and are not
time dependent. We provide explicit examples, and we stress the
difference between arbitrary gauge fixing and the construction of
observables which are indeed amenable to measurement.

We begin in section 2 with a technical presentation of the classical
``problem of time'' and with two familiar examples that exhibit
it, the relativistic free particle and conventional general
relativity. The reparameterization symmetry of the free particle
is nontrivial, and therefore this toy model offers edifying
illustrations of many ideas and techniques related to time
evolution, gauge fixing, and reparametization invariants. A
resolution of the time puzzle is given in  Section 3 where we
address and dismiss,  from a conceptual point of view,
 the supposed equivalence of time evolution
and gauge transformation. In
section 4 the problems associated with gauge fixing procedures are
analyzed and resolved. In section 5 we introduce Komar-Bergmann
intrinsic coordinates which make use of Weyl curvature scalars.
In section 6 we show that the Komar-Bergmann approach can be
intepreted as a gauge fixing procedure that fulfills the
requirements discussed in section 4, and we show in particular
that time dependence is necessarily retained through the
compulsory use of at least one explicitly time-dependent gauge
condition. Section 7 is devoted to the issue of observables. We
present with full detail the  well known result that scalar functions of
intrinsic coordinates which are themselves defined as scalar
functions of dynamical variables are diffeomorphism invariants.
The construction is carried out in complete detail for the free
particle where we confirm that the most general class of
invariants  are {\it not} constants of the motion. (Henceforth,
``invariant'' and ``gauge invariant'' will mean the same.) Then a
somewhat different perspective (with equivalent results) is
given for general relativity. We present our conclusions in
section 8, including possible implications of this work regarding
an eventual quantum theory of gravity.

%%%%%%%%%%%%%%%%%%%%%%%%%%%%%%%%%%%%%%%%%%%%%%%%%%%%%%%%%%%%%%%%%
%%%%%%%%%%%%%%%%%%%%%%%%%%%%%%%%%%%%%%%%%%%%%%%%%%%%%%%%%%%%%%%%%
\section{Generally covariant theories in phase space}
Here we review the formulation of generally covariant theories in
phase space with its diffeomorphism-induced gauge group. We also
consider the possibility that, besides diffeomorphism invariance,
internal gauge symmetries may be present, thus including cases
like Einstein-Yang-Mills, tetrad, and connection formulations of
general relativity 
\cite{pss:1997pr,pss:2000jmp,pss:2000grg,pss:2000pr}. Our starting
point is always a variational principle formulated with a
Lagrangian density, which is a function in
configuration-velocity space. Its corresponding phase space
formulation is given by the Dirac-Bergmann theory of constrained
dynamical systems.

The Dirac Hamiltonian takes the form \beq H_{\lambda} = n^A {\cal
H}_{A} + \lambda^A P_A ,  \label{HD} \eeq where $\lambda^A$ are
arbitrary functions of spacetime coordinates. The canonical variables $ n^A$ are fixed by
these arbitrary functions under time evolution and are often
called the gauge functions, although the redundancy of variables
that is caused by the gauge symmetry is not exhausted by them.
Their canonical momenta $P_A$ are primary constraints. The
physical phase space is further constrained by secondary
constraints ${\cal H}_A$. These constraints do not depend on $
n^A$ or $P_{A}$. If there is no symmetry in
configuration-velocity space beyond general covariance, the range
of the index $A$ is simply the dimensionality of the underlying
coordinate manifold and $n^0=n$ and $n^a$ are the lapse and shift variables. If additional internal symmetries are
present $A$ will also range over the dimension of the group Lie
algebra. This is the case for example with Yang-Mills gauge
fields in general relativity and also for tetrad connection approaches
to gravity.

The complete generator of infinitesimal gauge symmetries
which are projectable  onto phase space under the Legendre map takes the
general form
\begin{equation} G_{\xi }(t) = P_{A} \dot\xi^{A} + ({\cal H}_{A} +
P_{C''}n^{B'}{\cal C}^{C''}_{AB'})\xi^{A}\ , \label{GGGGG}
\end{equation}
where the structure functions are obtained from the
closed Poisson bracket algebra
\begin{equation} \{ {\cal H}_{A},{\cal
H}_{B'} \} = {\cal C}^{C''}_{AB'} {\cal H}_{C''}\ ,
\end{equation} and
where spatial integrations at time $t$ over corresponding repeated
capital indices are assumed hereafter. The generators $G_{\xi
}(t)$ act on phase space through the equal time Poisson
brackets, and map solution trajectories into other solutions. In
this sense, it is assumed that all phase
space variables appearing in (\ref{GGGGG}) are solution
trajectories $y_{\lambda}(t)$. Poisson brackets at time $t$ are
evaluated with respect to the canonical set $y_{\lambda}(t)$. The
``descriptors'' $\xi^A$ are arbitrary spacetime functions and
$\dot \xi^A$ stands for the time derivative of $\xi^A$. When
internal symmetries are present, the previously projectable
diffeomorphisms which alter spacetime foliations are no longer
projectable to phase space; they must be accompanied by \
internal gauge ``rotations'' fixed by the spacetime descriptors
$\xi^\mu$ \cite{pss:2000jmp,pss:2000grg,pss:2000pr}.

Notice that since dynamically $\dot n_{\lambda}^A(t)$ equals
$\lambda^A(t)$, when the functions $\xi^{A}$ take the values
$n^A$,  it appears that the gauge generator $G_{\xi }$ coincides
with the Dirac Hamiltonian (\ref{HD}). This is the technical setting
of the problem; $H_{\lambda}$ appears naively to be included
within the family of $G_{\xi}$, leading to the (spurious)
conclusion that the motion generated by $H_{\lambda}$ is gauge.

Now we present two examples of generally
covariant theories that exhibit the phenomenon just described.
%%%%%%%%%%%%%%%%%%%%%%%%%%%%%%%%%%%%%%%%%%%%%%%%%%%%%%%%%%%%%%%%%
%%%%%%%%%%%%%%%%%%%%%%%%%%%%%%%%%%%%%%%%%%%%%%%%%%%%%%%%%%%%%%%%%
\subsection{The relativistic free particle}
%%%%%%%%%%%%%%%%%%%%%%%%%%%%%%%%%%%%%%%%%%%%%%%%%%%%%%%%%%%%%%%%%
%%%%%%%%%%%%%%%%%%%%%%%%%%%%%%%%%%%%%%%%%%%%%%%%%%%%%%%%%%%%%%%%%

We employ the Lagrangian
$$ L = \frac{1}{ 2 n} \dot q^2 - \frac{1}{ 2} n \ ,
$$
where $q^\mu$ are the Cartesian spacetime coordinates for the
trajectory of a unit mass particle in Minkowski space and the
auxiliary variable $n$ plays the role of a lapse function on the
one-dimensional parameter space. The resulting Dirac Hamiltonian
is \beq H_{\lambda}(t)= \frac{1}{ 2} n ( p^2 + 1) + \lambda(t) \pi
\ , \label{fpham} \eeq where $ p_\mu $ and $\pi $ are the
variables conjugate to $ q^\mu$ and $n$. $\lambda$ is an
arbitrary function that reflects the reparametrization gauge
freedom of the model. Notice that the equations of motion yield
$\dot n = \lambda$. $n$ is therefore fixed, up to an integration
constant, by the arbitrarily chosen function $\lambda$. The gauge
generator $G_{\xi}(t)$ is constructed with the first class
constraints $(p^2 + 1) \approx 0$ and $\pi \approx 0$: \beq
G_{\xi}(t) = \frac{1}{ 2} \xi (p^2 + 1) + \dot\xi \pi \ ,
\label{gforfree} \eeq and $\xi$ is an arbitrary function. $G_{\xi
}(t)$ generates variations of dynamical variables resulting from
infinitesimal reparametrizations of the form $t' = t - n^{-1}(t)
\xi(t)$. Note that since the dynamics fixes $\dot n = \lambda$,
when $\xi$ happens to be equal to $n$  times an obvious infinitesimal factor, $H_{\lambda}$ is a
particular case of $G_{\xi}$ . Hence the claim, that we will prove
spurious, that dynamical evolution is a gauge transformation.
%%%%%%%%%%%%%%%%%%%%%%%%%%%%%%%%%%%%%%%%%%%%%%%%%%%%%%%%%%%%%%%%%
%%%%%%%%%%%%%%%%%%%%%%%%%%%%%%%%%%%%%%%%%%%%%%%%%%%%%%%%%%%%%%%%%
\subsection{Conventional canonical general relativity}
%%%%%%%%%%%%%%%%%%%%%%%%%%%%%%%%%%%%%%%%%%%%%%%%%%%%%%%%%%%%%%%%%
%%%%%%%%%%%%%%%%%%%%%%%%%%%%%%%%%%%%%%%%%%%%%%%%%%%%%%%%%%%%%%%%%

In
canonical general relativity the Dirac Hamiltonian takes the form
\begin{equation}
   H_{\lambda} = \left(n^\mu {\cal H}_\mu + \lambda^\mu P_\mu
\right) \ ,
\end{equation}
where (we use in the following the standard index notation
$\mu=(0,a)$)  $n^0 := n$ is the lapse and $n^a$ are the components of the
shift 3-vector . $P_\mu $ are the variables conjugate to the lapse
and shift and are primary constraints. ${\cal H}_\mu \approx 0$
are the so-called Hamiltonian and momentum secondary constraints.
The gauge generator is, at a given time $x^0$,
\begin{equation}
   G_{{\xi }}(x^0) = P_\mu \dot\xi^\mu + ( {\cal H}_\mu
+ n^\rho C^\nu_{\mu\rho} P_\nu) \xi^\mu . \label{thegen}
\end{equation}
(From now on, repeated index summation includes an
integration over 3-space.) The $\xi^\mu$ are arbitrary descriptor
functions of the spacetime coordinates and $C^\nu_{\mu\rho}$ are the
structure functions for the Poisson bracket algebra of the Hamiltonian
and momentum constraints. The functions $\xi^\mu$ are related to the
functions $\epsilon^\mu$ that define an infinitesimal diffeomorphism (in
the passive view: $x^\mu \rightarrow x^\mu - \epsilon^\mu$) by the
following construction \cite{pss:1997pr,berg-kom72}. Construct the
vectors ${\cal N}^\mu$ orthonormal to the constant-time surfaces,
\beq  {\cal N}^\mu
=\delta^\mu_0 n^{-1} - \delta^{\mu}_a n^{-1} n^a \ .
\label{lapseandshift}
\eeq
Then
\begin{equation}
   \epsilon^\mu = \delta^\mu_a \xi^a + {\cal N}^\mu \xi^0. \
\label{therecipy}
\end{equation}
Note that the Hamiltonian coincides
with the gauge generator when the arbitrary functions $\xi^\mu$ are
are chosen to be $\xi^\mu = n^\mu$.
Thus it appears at first sight that in this case the
gauge generator and Hamiltonian are identical. In
the next section we will present in depth arguments that there is no
problem of time if the roles of gauge operator versus the Hamiltonian
are properly understood.

%%%%%%%%%%%%%%%%%%%%%%%%%%%%%%%%%%%%%%%%%%%%%%%%%%%%%%%%%%%%%%%%%
%%%%%%%%%%%%%%%%%%%%%%%%%%%%%%%%%%%%%%%%%%%%%%%%%%%%%%%%%%%%%%%%%
\section{The resolution of the time evolution versus gauge problem}

\subsection{The space of field configurations}
%%%%%%%%%%%%%%%%%%%%%%%%%%%%%%%%%%%%%%%%%%%%%%%%%%%%%%%%%%%%%%%%%
%%%%%%%%%%%%%%%%%%%%%%%%%%%%%%%%%%%%%%%%%%%%%%%%%%%%%%%%%%%%%%%%%

The first answer we give to the question as to whether in
generally covariant theories the dynamical evolution is just
gauge follows from this consideration: gauge transformations, as
a special case of symmetries, map solutions of the equations of
motion into solutions. Therefore the natural arena for the action
of gauge transformations is just the space of solution field
configurations, i.e., the space of histories. In
reparametrization covariant particle models, for example, these
are just the particle world lines. An element in this space is a
specific space-time description --a history-- of the fields and
particles that are present in the physical setting. The action of
the gauge group on this space defines orbits. An orbit is the set
of all field configurations connected by diffeomorphisms. In the
passive view of diffeomorphisms, an orbit is understood as the
set of all field configurations that correspond to a unique
physical situation but expressed in different coordinates.
Infinitesimal variations of histories in the active point of view
are simply Lie derivatives along the direction of the vector
field $\epsilon^{\mu}$, associated with  infinitesimal coordinate
transformations of the passive view. In general relativity some
of these coordinate choices may have physical content in the
sense that each may correspond to a set of observers with a
scheme for physically achieving time simultaneity and
readjustment of proper time clocks. But the theory must also
carry with it instructions on how to move from one coordinate
fixing to another; this is the action of the gauge group. On the
other hand, the role of the Hamiltonian could not be more
distinct: it defines, through the Poisson brackets, the
differential equations that enable us to build the whole
configuration of the fields out of  initial data at a given
equal-time surface. It is obvious then that the Hamiltonian has no
action on the space of field configurations for it simply defines
how to build the elements of this space.
%%%%%%%%%%%%%%%%%%%%%%%%%%%%%%%%%%%%%%%%%%%%%%%%%%%%%%%%%%%%%%%%%
%%%%%%%%%%%%%%%%%%%%%%%%%%%%%%%%%%%%%%%%%%%%%%%%%%%%%%%%%%%%%%%%%
%%%%%%%%%%%%%%%%%%%%%%%%%%%%%%%%%%%%%%%%%%%%%%%%%%%%%%%%%%%%%%%%%
\subsection{Finite evolution and gauge operators}
%%%%%%%%%%%%%%%%%%%%%%%%%%%%%%%%%%%%%%%%%%%%%%%%%%%%%%%%%%%%%%%%%
%%%%%%%%%%%%%%%%%%%%%%%%%%%%%%%%%%%%%%%%%%%%%%%%%%%%%%%%%%%%%%%%%

The equations of motion fix the evolution of the gauge variables
after the arbitrary functions $\lambda^A$ have been selected.
We may then write down
a formal solutions of the dynamical equations, given initial
conditions at time $t = 0$, in terms of the finite evolution
operator
\beq U_{\lambda}(t) := T exp\left(\int_0^t dt'\{ - , \,
H_{\lambda}(t')\}_{y_0}\right), \label{tevolution}
\eeq
where $T$
stands for the $t$-ordering operator that places the highest $t$
on the right. All Poisson brackets in the expansion defined on
the right hand side are evaluated in terms of $y_0 := y(t=0)$. (We
must only be careful to take account of the explicit time
dependence of the functions $\lambda^A$). Thus it is possible to
express all dynamical variables, including the arbitrary gauge
variables $n^A$ in terms of initial values $y_0$.

The finite form of the gauge transformation looks quite different:
\beq V_{\xi}(s,t) =  exp\left(s  \{ - , \,
G_{\xi }(t)\}_{y_{\lambda}(t)} \right) \ . \label{gevolution}
\eeq
(The functions $\xi$, being arbitrary, may develop a dependence
on the parameter $s$; in this case the finite operator for gauge
transformations will contain an $s$-ordering operator as well
\cite{ps04}.)

%%%%%%
%%%%%%%%%%%%%%%%%%%%%%%%%%%%%%%%%%%%%%%%%%%%%%%%%%%%%%%%%%%%%%%%%%%%
\subsection{Application to the relativistic free particle}
%%%%%%%%%%%%%%%%%%%%%%%%%%%%%%%%%%%%%%%%%%%%%%%%%%%%%%%%%%%%%%%%%%%%%%%
%%%%%%%%%%%%%%%%%%%%%%%%%%%%%%%%%%%%%%%%%%%%%%%%%%%%%%%%%%%%%%%%%%%%%%%

We demonstrate the action of (\ref{tevolution}) with the free
particle using (\ref{fpham}). We find
\bea
n(t) &=& T exp\left(\int_0^t
dt'\{ - ,H_{\lambda}(t')\}_{y}\right) n_0 = n_0 +
\int_0^t dt_1\{ n_0
, H_{\lambda}(t_1)\} \\&=& n_0 + \int_0^t dt_1\{ n_0
,\lambda(t_1)\pi \} = n_0 + \int_0^t dt_1\lambda(t_1). \label{Nt}
\eea
(In this expression and henceforth we will let the variable name with the zero subscript represent the initial value of the variable).  Notice
that all the remaining nested Poisson brackets vanish since the
first yields a numerical function. Similarly, $p_\mu (t) = p_\mu$, and
\bea q^\mu(t) &=& q^\mu_0 +
\int_0^t dt_1\{ q^\mu_0 ,\frac{n_0}{2} p^2\} + \int_0^t dt_2
\int_0^{t_2} dt_1\{\{ q^\mu_0 ,\frac{n_0 }{2} p^2\},\lambda(t_1)\pi_0\}
\nonumber \\ &=& q^\mu + \int_0^t dt_1 n_0 p^\mu + \int_0^t dt_2
\int_0^{t_2} dt_1 p^\mu \lambda(t_1) \nonumber \\ &=& q^\mu_0 +
p^\mu \int_0^t dt_1 n_0(t_1). \label{qt} \label{thetraj}
\eea
Since we shall require this result below when we compute the
action of the finite gauge generator,
 let us also calculate the evolution of the constraint
 $\pi(t)$,
\beq
\pi(t) = \pi_0+ t \{\pi_0 , \frac{1}{ 2} N (p^{2}+1)\} =
\pi_0 -\frac{1}{ 2}
(p^{2}+1) t. \label{pit}
\eeq
(Note that the additional constraint term is required to preserve the
canonical Poisson Bracket $\{\pi(t),q^{\mu}(t)\} = 0$.)

Let us return to the generator of gauge transformations
$G_{\xi}(t)$. For each time $t$ this object
generates an infinitesimal variation of solution trajectories to
produce new solution trajectories.

In an effort to minimize misunderstandings concerning the action
of the finite gauge generator (\ref{gevolution}) we will
calculate its action in two equivalent ways, first calculating
Poisson brackets with respect to the canonical variables
$y(t)$, and then alternatively in terms of the initial variables $y_0$.

First, using the gauge generator (\ref{gforfree}), written as
$G_{\xi} = \frac{1}{ 2} \xi(t) (p^{2}(t) + 1) + \dot \xi(t)
\pi(t))$, we find
\bea
q_ s^{\mu}(t) &=& q^{\mu}(t) + s \frac{\partial G_{\xi}(t)
}{ \partial p_{\mu}(t) } \nonumber \\
&=& q^{\mu}(t) + s \xi(t) p^\mu (t) = q^{\mu}(t) + s \xi(t) p^\mu. \eea Of course, since the
$y(t)$ are obtained from $y$ through a canonical
transformation, we can equivalently calculate Poisson brackets
with respect to the initial variables $y_0$:
\bea q^\mu_{s}(t) &=& \left(exp(s  \{ - , \, G_{\xi}(t) \}_{y_0})\right)
q^\mu(t)
\nonumber \\
&=&q^{\mu}(t)+ s \{ q^{\mu}_0 +p^{\mu} n_0 t  , \, \frac{1}{ 2}\xi(t))
p^{2}
+\dot \xi(t)(\pi_0 -\frac{1}{ 2} t (p^{2} + 1) \}_{y_0} \nonumber \\
&=& q^\mu(t) + s \xi(t) p^\mu \ , \label{thetrajtrans}
\eea

The corresponding expression for $n_{s}(t)$
is
\beq
n_{s}(t) =
\left(exp(s  \{ - , \, G_{\xi}(t) \}_{y_{\lambda}(t)}
\right)n(t) = n(t) + s \dot
\xi(t) \ . \eeq

%%%%%%%%%%%%%%%%%%%%%%%%%%%%%%%%%%%%%%%%%%%%%%%%%%%%%%%%%%%%%%%%%
%%%%%%%%%%%%%%%%%%%%%%%%%%%%%%%%%%%%%%%%%%%%%%%%%%%%%%%%%%%%%%%%%
\section{The
gauge fixing resolution of the evolution versus gauge puzzle}
%%%%%%%%%%%%%%%%%%%%%%%%%%%%%%%%%%%%%%%%%%%%%%%%%%%%%%%%%%%%%%%%%
%%%%%%%%%%%%%%%%%%%%%%%%%%%%%%%%%%%%%%%%%%%%%%%%%%%%%%%%%%%%%%%%%
Another way
to rephrase the claim ``dynamical evolution equals gauge transformation''
makes use of the gauge fixing (GF) methods. For, it is argued, suppose we
consider a complete set of GF constraints, say $\chi^{A} = 0 $,
complete in the
sense that they eliminate all of the gauge freedom:
\begin{equation} \{ \chi^{A} , \, G_{\xi}(t) \}= 0, \ \
\forall t  \Rightarrow \xi^{A} =0 \ ,
\label{frozen}
\end{equation}
that is, the arbitrary functions $\xi^{A}$ in $G_{\xi}(t)$ (see
equation \ref{GGGGG}) become zero and no gauge freedom is left.
Equation (\ref{frozen}) expresses the fact that, after
implementing the GF constraints, the gauge evolution is frozen
because any gauge motion will take our field configurations (or
trajectories) out of the GF constraints surface. Then one makes
the assertion: since $H$ is a particular case of $G(t)$, the
dynamics must therefore also be frozen, for the dynamics will
take the field configurations out of the GF constraints surface.
But what might seem an insurmountable problem is easily overcome
when we recognize that the GF constraints (or at least one of
them) may depend on the time variable. The dynamical evolution
for an explicitly time dependent constraint is
$$ \frac{d \chi }{dt} = \frac{\partial \chi}{
\partial t} + \{ \chi , \, H \} \ ,$$
and to require this to vanish no longer freezes the dynamics. In
fact what seemed to be a problem is a theorem: {\sl in generally
covariant theories, at least one of the GF constraints must
exhibit an explicit dependence on time}
\cite{pss:1997pr,ps:95cqg}. This time dependent constraint plays
the role of defining the time in terms of the dynamical
variables. This argument will be worked out in full detail for
the free relativistic particle and for general relativity.
%%%%%%%%%%%%%%%%%%%%%%%%%%%%%%%%%%%%%%%%%%%%%%%%%%%%%%%%%%%%%%%%%
%%%%%%%%%%%%%%%%%%%%%%%%%%%%%%%%%%%%%%%%%%%%%%%%%%%%%%%%%%%%%%%%%
%%%%%%%%%%%%%%%%%%%%%%%%%%%%%%%%%%%%%%%%%%%%%%%%%%%%%%%%%%%%%%%%%
%%%%%%%%%%%%%%%%%%%%%%%%%%%%%%%%%%%%%%%%%%%%%%%%%%%%%%%%%%%%%%%%%
\subsection{The Dirac bracket puzzle}
%%%%%%%%%%%%%%%%%%%%%%%%%%%%%%%%%%%%%%%%%%%%%%%%%%%%%%%%%%%%%%%%%
%%%%%%%%%%%%%%%%%%%%%%%%%%%%%%%%%%%%%%%%%%%%%%%%%%%%%%%%%%%%%%%%%
The Dirac bracket argument for frozen dynamics is the
previous GF argument in disguise. It proceeds as
follows. The GF procedure makes the first class constraints of
the theory second class through the addition of the appropriate GF
constraints. Since the Dirac Hamiltonian $H_\lambda$ in a generally covariant
theory is made up of first class constraints, when the Dirac
bracket $\{\, - , -  \}^*$ is introduced, all constraints (which are now second class) can
be taken to be zero inside the bracket, that is
$$
\{ -, \, H_\lambda \}^* = \{ -, \, 0
\}^* = 0 \ .
$$
Then it appears that no dynamical evolution remains,
independently of whether the GF constraints have an explicit
time dependence or not. As before
the flaw in this argument can be traced back to a failure to
appropriately take into
account the presence of time dependent gauge fixing constraints.
Our starting
point is a first class Dirac Hamiltonian
$$
H_{\lambda} = H_c + \lambda^i \phi_i \ ,
$$
where $H_{c}$ is the canonical Hamiltonian and the $\phi_i$ are
first class constraints. We implement a set $\chi_i$ of GF
constraint. The on-shell dynamics does not change if we
substitute all constraints, the original and the gauge fixing
constraints, into the Dirac Hamiltonian, each multiplied by a
different arbitrary function that plays the role of a Lagrange
multiplier. The stabilization of the constraints, i.e., their
conservation in time, then determines all these Lagrange
multipliers. Let us use the notation $\psi_n$ for the complete set of
now second class constraints. Using the extended Hamiltonian
$H_{e\,\lambda} = H_c + \lambda^n \psi_n$ the   result on
shell is \beq
\frac{d}{ dt}\psi_n = \frac{\partial
}{
\partial t}\psi_n + \{ \psi_n , \, H_c \} + \lambda^m \{ \psi_n ,
\,\psi_m \} = 0 \ ,
\eeq
 that determines
\beq \lambda^m =
-\big(\frac{\partial }{
\partial t}\psi_n + \{ \psi_n , \, H_c \}\big) M^{mn} \ ,
\eeq
with $ M^{mn}$ being the inverse matrix of $\{ \psi_m , \,\psi_n
\}$ . Substituting these values for $\lambda^m$ into
$H_{e\,\lambda}$ we obtain (always on the constraint surface)
\beq
\{-, \, H_{e\,\lambda} \} = \{-, \, H_c \}^* -\{-,
\, \psi_m \}M^{mn}\frac{\partial \psi_n}{
\partial t} \ ,
\label{gfsolution}
\eeq
where we have used the standard notation for the Dirac brackets,
$$
\{-, \, - \}^* = \{-, \, - \} - \{-, \, \psi_m \}M^{mn}\{\psi_n, \, - \} \ .
$$
Therefore, even when the first term on the right hand side in
(\ref{gfsolution}) vanishes (as is the case in all generally
covariant theories), a nontrivial dynamical evolution still
obtains as long as at least one GF constraint has an explicit
time dependence, where the gauge-fixed Hamiltonian is
\beq
H_{GF} = \psi_m M^{mn} \frac{\partial \psi_n}{\partial t}. \label{hgf}
\eeq

\subsection{Dirac brackets for the free relativistic particle}

We shall illustrate the nontrivial evolution which results from an
explicitly time-dependent gauge fixing constraint with the free
relativistic particle. Let us impose the gauge condition $\psi_{3}:=
q^{0}(t)-f(t) \approx 0$, where $f$ is an arbitrary monotonically
increasing function. Letting $\psi_{1} := \frac{1}{ 2} (p^{2} + 1)
\approx 0$ and $\psi_{2} := \pi \approx 0$ represent the original
first class constraints, preservation of the constraint $\psi_{3}$
under time evolution leads to a fourth constraint $\psi_{4} := p^{0}
n(t) - \dot f(t) \approx 0$. The Poisson brackets among the constraints
displayed as a matrix then takes the form
\beq
\{ \psi_{m}, \psi_{n} \} = \left( \begin{array}{cccc}
0 & 0 & -p^{0} & 0 \\
0 & 0 & 0 & -p^{0} \\
p^{0} & 0 & 0 & -n(t) \\
0 & p^{0} & n(t) & 0 \end{array} \right),
\eeq
with inverse
\beq
M^{mn} = \left( \begin{array}{cccc}
0 & -\frac{n(t) }{ (p^{0})^{2}} & \frac{1}{ p^{0}} & 0 \\
\frac{n(t) }{ (p^{0})^{2}} & 0 & 0 & \frac{1}{ p^{0}} \\
-\frac{1 }{ p^{0}} & 0 & 0 &0 \\
0 & -\frac{1}{ p^{0}} &0 & 0 \end{array} \right). \label{Mmn} \eeq Since the
non-vanishing explicit time derivatives of the constraints are $
\frac{\partial \psi_3}{ \partial t} = \dot f$ and $\frac{\partial
\psi_4}{ \partial t} = \ddot f$, the extended Dirac
Hamiltonian, once we have set the canonical Hamiltonian to
zero inside the Dirac bracket, see (\ref{hgf}), becomes
\beq
H_{GF} = \psi_{1}M^{13} \dot f + \psi_{2}M^{24} \ddot f
= \frac{1}{ 2 p^{0} }(p^{2} +1) \dot f + \frac{1}{ p^{0} } \pi
\ddot f. \label{fham} \eeq This yields the equations of motion
\beq \dot q^{\mu}(t) = \frac{p^{\mu}}{ p^{0}} \dot f(t), \label{qmudot}\eeq and \beq
\dot n(t) = \frac{\ddot f(t) }{ p^{0}}. \label{dotN}
\eeq

%%%%%%%%%%%%%%%%%%%%%%%%%%%%%%%%%%%%%%%%%%%%%%%%%%%%%%%%%%%%%%%%%
%%%%%%%%%%%%%%%%%%%%%%%%%%%%%%%%%%%%%%%%%%%%%%%%%%%%%%%%%%%%%%%%%

\section{Intrinsic coordinates in generally covariant theories}
%%%%%%%%%%%%%%%%%%%%%%%%%%%%%%%%%%%%%%%%%%%%%%%%%%%%%%%%%%%%%%%%%
%%%%%%%%%%%%%%%%%%%%%%%%%%%%%%%%%%%%%%%%%%%%%%%%%%%%%%%%%%%%%%%%%

In this and the following section we introduce the gauge fixing
method of Komar and Bergmann  that has a direct application to
the preceding discussion. The method employs Weyl scalars to fix
intrinsic coordinates. We first present their definition, and
show that they do not depend on the lapse and shift. Komar and Bergmann proposed their use in
vacuum spacetimes. We shall show that they can equally well be
used in spacetimes with other fields present, as we will see in
the Einstein-Maxwell case. In the following subsection we show
that regardless of the arbitrary coordinate system in which one
may be working initially, transformation to the intrinsic
coordinate system yields identical metric functions. The
explanation of the use of Weyl scalars as a gauge fixing is given
in section 6.

\subsection{Weyl curvature scalars}

We begin with the general expression for the conformal tensor in
terms of the Riemann tensor,
\beq
C_{\mu \nu \rho \sigma} = R_{\mu \nu \rho \sigma} - g_{\mu [\rho}
R_{\sigma ] \nu} + g_{\nu [\sigma} R_{\rho ] \mu} + \frac{1}{ 3} R
g_{\mu [\rho} g_{\sigma [\nu}. \label{cr}
\eeq
We will be concerned only with the conformal tensor evaluated on
solutions of the equations of motion,
\beq
R_{\mu \nu} = 8\pi (T_{\mu \nu} - \frac{1}{2} g_{\mu \nu} T),
\eeq
where $T_{\mu \nu}$ is the stress-energy tensor and $T$ is its trace.

In the vacuum case which we consider initially where $R_{\mu \nu} := R_{\rho
\mu \nu \sigma} g^{\rho \sigma} = 0$ the conformal and Riemann tensors
coincide on shell.  Bergmann and Komar discovered that spatial
components of the Riemann tensor, and also contractions with the
normal ${\cal N}^{\mu}$ to the fixed time hypersurfaces could be expressed in
terms of canonical variables \cite{berg-kom60,berg-book}.
The construction uses the projection tensor
\beq
e^{\mu \nu} := g^{\mu \nu} + {\cal N}^{\mu} {\cal N}^{\nu}, \label{e}
\eeq
and the Gauss-Codazzi relations,
\beq
R_{abcd} = {}^{3}R_{abcd} - K_{bc} K_{ad} + K_{bd} K_{ac}, \label{gc1}
\eeq
and
\beq
R_{\mu abc} {\cal N}^{\mu} = K_{ab|c} - K_{ac|b}, \label{gc2}
\eeq
with the observation that the canonical momentum written in terms of
the extrinsic curvature $K_{ab}$ is
\beq
\pi_{ab} = \sqrt{g} ( K_{ab} - K^{c}_{c} g_{ab}).
\eeq
Thus we may invert to find the extrinsic curvature in terms of the
momentum
\beq
K_{ab} = \frac{1}{ \sqrt{g}} ( \pi_{ab} - \frac{1}{ 2} \pi^{c}_{c} g_{ab}).
\label{kab}
\eeq
In all of these expressions indices are raised with $e^{ab}$
and ``$|$'' signifies covariant derivative with respect to the
spatial metric.

Referring to (\ref{gc1}) and (\ref{kab}) we see that on shell the
spatial components of the conformal tensor may be written in terms of
canonical variables as promised,
\beq
C_{abcd} = {}^{3}R_{abcd} - K_{bc} K_{ad} + K_{bd} K_{ac}.
\eeq
From (\ref{gc2}) we have on shell
\beq
C_{abc}:=C_{\mu abc} {\cal N}^{\mu} = K_{ab|c} - K_{ac|b}.
\eeq
Finally we find on shell, using (\ref{e}), that
\beq
C_{ab}:= C_{\mu a b \nu} {\cal N}^{\mu} {\cal N}^{\nu} = C_{c a b d} e^{cd}.
\eeq
All three expressions can therefore independent of the canonical variables ${\cal N}$ and ${\cal N}^a$.

We are finally prepared to construct the Weyl scalars which are most
conveniently written for our purposes as \cite{berg-book}
\bea
W^{1} &=& C_{\alpha \alpha' \beta \beta' } g^{\beta \beta'
\gamma \gamma'}
C_{\gamma \gamma' \delta \delta' } g^{\delta \delta' \alpha \alpha'},
\label{a1} \\
W^{2} &=& C_{\alpha \alpha' \beta \beta' } g^{\beta \beta' \gamma \gamma'}
C_{\gamma \gamma' \delta \delta' } \epsilon^{\delta \delta' \alpha
\alpha'}, \\
W^{3} &=& C_{\alpha \alpha' \beta \beta' } g^{\beta \beta' \gamma \gamma'}
C_{\gamma \gamma' \delta \delta' } g^{\delta \delta' \rho \rho'}
C_{\rho \rho' \sigma \sigma' } g^{\sigma \sigma' \alpha \alpha'}, \\
W^{4} &=& C_{\alpha \alpha' \beta \beta' } g^{\beta \beta' \gamma \gamma'}
C_{\gamma \gamma' \delta \delta' } g^{\delta \delta' \rho \rho'}
C_{\rho \rho' \sigma \sigma' } \epsilon^{\sigma \sigma'\alpha \alpha'},
\eea
where
\bea
g^{\beta \beta' \gamma \gamma'} &:=& 2 g^{\beta [\gamma} g^{\gamma'] \beta'}
\nonumber \\
&=& 2 e^{\beta [\gamma} e^{\gamma'] \beta'} - 2 {\cal N}^{\beta} {\cal N}^{[\gamma}
e^{\gamma'] \beta'} - 2 e^{\beta [\gamma} {\cal N}^{\gamma' ]}{\cal N}^{\beta'}.
\label{gabcd}
\eea
Substitution of (\ref{gabcd}) into (\ref{a1}) yields, for example
\cite{berg-kom60},
\beq
W^{1} = C_{abcd} C^{abcd} +4 C_{abc} C^{abc} + 4 C_{ab} C^{ab}.
\eeq

One might suspect from (\ref{cr}) that this construction could be
generalized to include other fields. We shall now show that after
substitution of the Einstein equations into (\ref{cr}) the right
hand side is indeed independent of the lapse and shift, and
depends only  on the remaining material and  metric phase space
variables. We carry out the construction explicitly for
Einstein-Maxwell theory.

The stress-energy tensor is up to a constant
$$
T_{\mu \nu} = F_{\alpha \mu} F_{\beta \nu} g^{\alpha \beta} -
\frac{1}{4} g_{\mu \nu} F_{ \alpha \beta} F^{ \alpha \beta},
$$
where $F_{\alpha \beta}$ is the Maxwell tensor.
We need the spatial components, and we want to write them in terms of
the canonical momentum
$$
P^a = \sqrt{|{}^4 g|} F_{\mu \nu} g^{a \mu} g^{0 \nu}
$$
Substituting for the metric it turns out that
$$
P^a = \sqrt{|{}^4 g|} n^{-2} (F_{0 b} e^{a b} + F_{b c} e^{a
b} n^c ),
$$
so
$$
F_{0 a} = \frac{1}{\sqrt{|{}^4 g|}} n^2 P_a - F_{ab}n^b,
$$
where the index is lowered with the three-metric.
Let's first substitute into the $ F_{ \alpha \beta} F^{ \alpha
\beta}$ term:
$$
F_{ \alpha \beta} F^{ \alpha \beta} = 2 F_{0a} F_{0b} g^{00}
g^{ab}
+2 F_{0a} F_{b 0} g^{0 b} g^{a 0} + 4 F_{0a} F_{b c} g^{0 b} g^{a c}
+ F_{ab}F_{cd } g^{a c} g^{ b d}.
$$
Some remarkable cancelations ultimately yield
$$
F_{ \alpha \beta} F^{ \alpha \beta} = - 2\frac{1}{|{}^4 g|}
n^2 P^a P^b g_{ab}
+ F_{ab} F_{cd} e^{ac} e^{bd}.
$$
A similar calculation yields
$$
F_{\alpha a} F_{\beta b} g^{\alpha \beta} = - \frac{1}{ |{}^4 g|} n^2 P_a P_b
+ F_{c a} F_{d b} e^{c d}.
$$
Noting that $|{}^4 g| = n g$, where $g$ represents the determinant
of the three metric, we find
$$
T_{a b} = - \frac{1}{ g} P_a P_b + \frac{1}{ 2 g} g_{a b} P^c P^d g_{cd}
- \frac{ 1}{ 4} F_{ab} F_{cd} e^{ac} e^{bd}+ F_{c a} F_{d b} e^{c
d} .
$$
Now we can calculate the required components and projections of the
conformal tensor. Start with (\ref{cr}), noting that for Einstein-Maxwell theory the
stress-energy
is traceless implying that the curvature scalar $R$ vanishes,
$$
C_{a b c d} = R_{a b c d} - g_{a [c} R_{d] b} + g_{b [c} R_{d]a}.
$$
So we merely
have to replace the Ricci tensor $R_{a b}$ by $8 \pi T_{a b} $,
and we have our first result, the independence of $C_{a b c d}$ of the
lapse and shift as promised.

The independence of $C_{a b c} := C_{\mu a b c} {\cal N}^\mu$ and
$C_{a b } :=
C_{\mu a b \nu} {\cal N}^\mu {\cal N}^\nu $ of the lapse and shift is simpler to see.
Consider first the contraction
$ ( - g_{\mu [a } R_{b ] \nu} + g_{\nu [a} R_{b ] \mu} ) {\cal N}^\mu$. The
first terms gives the covariant spatial component of the normal,
which is zero. The second term is a projection of the Einstein equations
(since $R = 0$), and is therefore a constraint. So both contributions
vanish on shell. A similar argument holds for the second projection.
So we have the final result: The Weyl scalars are independent of the lapse and shift canonical variables.

\subsection{Komar-Bergmann intrinsic coordinates in general relativity}
%%%%%%%%%%%%%%%%%%%%%%%%%%%%%%%%%%%%%%%%%%%%%%%%%%%%%%%%%%%%%%%%

We now show how the Weyl scalars can be used to construct
quantities which are invariant under diffeomorphisms.
Bergmann and Komar were pioneers in
this approach \cite{berg-kom60,komar58,berg61a,berg61b}.

We will consider only the
generic asymmetric case when the four Weyl scalars $W^I ; \ I =
0, \cdots
3$ are independent.

If the metric $g$ is locally
described in a given system of coordinates $\{ x^\mu \}$ as
$g^{\mu\nu}(x)$ then  four independent functions $A^{I}(W)$ of the
four scalars become four scalar functions of
the coordinates
\beq
a^I_g(x) :=  A^{I}(W[g(x)])  .
\label{ai}
\eeq
Independence of the four Weyl scalars and the functions $A^{I}$
 implies
\beq
\det(\frac{\partial a^I_g(x)}{ \partial x^\mu}) \neq 0.
\label{det}
\eeq

Consider a metric $g'$, infinitesimally close to
$g$, and related to it by an active diffeomorphism generated by the
infinitesimal vector field $\epsilon^\mu \partial_\mu$, with
$\epsilon^\mu(x)$ arbitrary functions (so $g$ and $g'$ belong to the
same gauge orbit). If $g'^{\mu\nu}(x)$ is the local description of $g$
in the coordinates $\{ x^\mu \}$, we can write
$$
{g'}^{\mu\nu}(x) =
g^{\mu\nu}(x) + {\cal L}_{\epsilon^\mu \partial_\mu} (g^{\mu\nu}(x)),
$$
where ${\cal L}_{\epsilon^\mu \partial_\mu}$ represents the Lie
derivative with respect to the vector $\epsilon^\mu \partial_\mu$.

Since $A^I$ are
scalars,
$$
A^I(W[g'(x')]) = A^I(W[g(x)]),
$$
that is,
\beq
a^I_{g'}(x') = a^I_g(x).
\label{ascalar}
\eeq
The Bergmann and Komar procedure consists
in implementing a metric-dependent change of coordinates dictated by the
functions
$a^I_g$. The new coordinates will be written as $X^I$, and are related
to the old ones by
\beq
X^I(x) := a^I_g(x).
\label{scalarcoord}
\eeq
(In section 6 we will find conditions which must be satisfied by the
functions $A^{I}$ in order that $X^{0}$ actually labels spacelike
foliations of spacetime.)
We will call these new coordinates ``intrinsic''. As is clear from the
notation, the change of coordinates (\ref{scalarcoord}) is
$g$-dependent. 

Consider now the passive coordinate transformation that results from 
first transforming from $x$ to $x'$, and then to intrinsic coodinates,
$X'(x'(x))$
$$
X'^I(x'(x)) := a^I_{g'}(x'(x)),
$$
Recalling (\ref{ascalar}), we find
$$ X'^I(x') =
X^I(x).
$$
Now we can express the metric $G$, used to define
the intrinsic coordinates $X^I$, in terms of these coordinates. It will
take the form $G^{IJ} (X)$, with
$$ G^{IJ} (X) = g^{\mu\nu}(x)
\frac{\partial X^I }{ \partial x^\mu} \frac{\partial X^J}{ \partial
x^\nu}.
$$
Notice that indices $I,J$, used to enumerate the four scalar
$A^I$,
now play a role indistinguishable from spacetime indices. Had we started
the whole procedure from $g'$ instead of $g$ we would have ended up with
$G'^{IJ} (X')$ instead of $G^{IJ} (X)$. But the fact
that the
new coordinates have been constructed out of scalars guarantees that the
functions $G'^{IJ}$ and $G^{IJ}$ coincide, as we
now demonstrate. Since $X'(x') = X(x)$,
\bea
G'^{IJ} (X(x))
&=& G'^{IJ} (X'(x')) =
g'^{\mu\nu}(x')\frac{\partial X'^I(x')}{ \partial x'^\mu}
\frac{\partial X'^J(x')}{ \partial x'^\nu} \nonumber \\
&=& g'^{\mu\nu}(x')
\frac{\partial X^I(x) }{ \partial x^\rho} \frac{\partial x^\rho}{\partial
x'^\mu} \frac{\partial X^J(x)}{ \partial x^\sigma} \frac{\partial x^\sigma
}{ \partial x'^\nu} \nonumber \\
&=& g^{\rho\sigma}(x) \frac{\partial X^I(x)}{
\partial x^\rho} \frac{\partial X^J(x) }{\partial x^\sigma} = G^{IJ}
(X(x)).
\label{samefunct1}
\eea

Let us recapitulate. We assume an observer has made an initial
arbitrary choice of coordinates $x$ and is working with a fixed
solution trajectory $g(x)$ in these coordinates. This first observer is
instructed how to select a new coordinate choice $X(x)$
resulting in a new functional form of solution $G(X)$ in
terms of the new coordinates. The metric at $x$ is mapped to a
metric at $X$. We assume that a second observer is working with the 
same physical solution trajectory, which means that the second
trajectory must be obtainable from the first through a passive
coordinate transformation $x'(x)$ with the functional form $g'(x')$. 
This second observer then follows the same instructions to transform to
intrinsic coordinates. We discover that the composite coordinate
transformation $X'(x'(x))$ is the same as the original one step
transformation $X(x)$. Thus the second observer agrees not only with
the functional form of the solution in intrinsic coordinates, but
also with the values of the individual metric components assigned to 
the same intrinsic coordinate location.

This procedure only works provided the intrinsic coordinates are
determined by scalar functions of the dynamical fields. Indeed, as we
shall discuss further in section 7, one can view these scalars as
identifying points in the spacetime manifold. In fact, this procedure
can be used to ``test'' \cite{komar58} the physical equivalency of a pair of
solutions. 
This will be the case if the metric tensors are
connected by a diffeomorphism.
The test is performed by going to the intrinsic
coordinates (\ref{scalarcoord}), and then checking whether the new
functions $G^{IJ}$ that describe the metric in the new coordinates
coincide. Notice that the choice of the intrinsic coordinates is subject
to the choice of the scalars $A^I$. Obviously, any ``coordinate
transformation'', subject to the conditions to be determined in
section 6,
\beq
A^I \rightarrow \hat A^I = f^I (A), \label{hata}
\eeq
defined by a set of functions $f^I$ (defining an invertible
coordinate transformation) gives a new set of scalars $\hat
A^I$ as good
as the former one as regards the application of our procedure.
Therefore, in order to correctly perform the test, the form of the
scalars $A^I$ must be given and agreed upon by all the observers
who wish to check whether their respective physics agrees.
%%%%%%%%%%%%%%%%%%%%%%%%%%%%%%%%%%%%%%%%%%%%%%%%%%%%%%%%%%%%%%%%%%%%%%%%%
%%%%%%%%%%%%%%%%%%%%%%%%%%%%%%%%%%%%%%%%%%%%%%%%%%%%%%%%%%%%%%%%%%%%%%
\subsection{Intrinsic coordinates for the relativistic free particle}
%%%%%%%%%%%%%%%%%%%%%%%%%%%%%%%%%%%%%%%%%%%%%%%%%%%%%%%%%%%%%%%%%%%%%%%
%%%%%%%%%%%%%%%%%%%%%%%%%%%%%%%%%%%%%%%%%%%%%%%%%%%%%%%%%%%%%%%%%%%%%%%

We now refer to the relativistic free particle for a simple implementation
of intrinsic coordinates. There are several ways to consider analogs of
the Weyl scalars for the relativistic free particle. One could use, for
instance, functions of the temporal coordinate $q^0$ of the
trajectory in Minkowski
space. The particle proper time is also a useful
analogue of the Weyl scalars. We will explore both cases.

Let us first use Minkowski time. In
this case we select the analogue of the Weyl scalar to be $A[q^\mu] =
f^{-1}(q^0)$, where $f^{-1}$ is an arbitrary monotonically increasing
function of its argument.  (Recall  that each component $q^{\mu}$
transforms as a scalar under reparametrizations.) We represent the
parameter time by $t$  and we set the intrinsic coordinate time $T$
equal to the appropriate scalar function of the dynamical variables
\beq
T:= a_q(t) := A[q(t)] = f^{-1}(q^0(t)). \label{theef}
\eeq

Referring to the general solution (\ref{thetraj}) our observer is instructed to
set
\beq
f(T) = q^{0}(t) = q^{0} + p^{0} \int_{0}^{t} dt_{1} n(t_{1}),
\eeq
so
\beq
\int_{0}^{t} dt_{1} n(t_{1}) = \frac{1}{ p^{0}} (f(T) - q^{0}).
\eeq

Also, differentiating ({\ref{theef}) we find
\beq
\frac{dt}{ dT} = \frac{\frac{df }{ d\bar t}}{ p^{0} n(t)}.
\eeq 
Substituting into the solution (\ref{thetraj})
 we find
\beq
 Q^\mu(T) :=
 q^{\mu}(t(T)) = q^{\mu}_0 +
\frac{p^{\mu} }{ p^{0}} (f(T) - q^{0}_0),
\label{qumut}
\eeq
and
\beq
 N(T) = n(t) \frac{dt}{dT} = \frac{\frac{df(T) }{ dT}}{ p^{0}} \label{barNt}
\eeq

 We want to compare this solution in terms of the
intrinsic coordinate $T$ to that obtained by a second observer working with a
different parametrization. Let us use our canonically implemented
reparametrization transformed solution (\ref{thetrajtrans}) since we wish to
analyze this construction later on from the perspective of gauge
fixing by transforming along gauge orbits. So our second observer is
instructed to set his $q^{0}_{s}(t_{s}) = f(T_{s})$, resulting in the
following determination of $f(T_{s})$ in terms of the gauge descriptor and
lapse,
\beq
s \xi (t_{s}) + \int_{0}^{t_{s}} dt_{1} n(t_{1}) = \frac{1}{ p^{0}}
(f(T_{s}) - q^{0}).
\eeq
Substituting into the gauge transformed  trajectory we find
again that
\beq
Q^{\mu}_{s}(T_{s}) :=
q^{\mu}_{s}(t_{s}(T_{s})) = q^{\mu}_0 + \frac{p^{\mu} }{ p^{0}}
(f(T_{s}) - q^{0}_0),
\eeq
and
\beq
 N_{s}(T_{s}) =  \frac{\frac{df (T_{s})}{ dT_{s}} }{ p^{0}}
\eeq

Thus $Q^{\mu}_{s}(T) =
Q^{\mu}(T) $  and $ N_{s}(T) = N(T) $
consistently with our general claim that all observers find the same solution.

Let us also examine what happens to the transformation to intrinsic
coordinates when we first go to coordinate $t'$ before passing to
intrinsic coordinates. For this purpose we will assume that we
undertake an infinitesmal coordinate transformation
$$
t' = t - n^{-1}(t) \xi (t),
$$
for infinitesmal $\xi (t)$, so
$$
t = t' + n^{-1}(t') \xi (t').
$$
Under this transformation we find
$$
q'^{0}(t') = q^{0}(t(t')) = q^{0}(t') + \dot q^{0}(t') n^{-1}(t')
\xi (t'),
$$
and therefore in passing to intrinsic coordinates
$$
T_{\xi}(t') = q^{0}(t(t')) = q^{0}(t') + \dot q^{0}(t') n^{-1}(t')
\xi (t') = q^{0}(t) =T (t).
$$
Of course we could have avoided writing this out in detail by using
the fact that $q^{0}$ transforms as a scalar under $t'(t)$.
Nevertheless it is instructive to see that even though the functional
dependence of $q^{0}$ does change we still find that $T_{\xi}(t'(t)) = T(t)$, i.e., it is the same transformation
from $t$ to $T$!

Proper time (and functions thereof) may also be used as intrinsic
coordinates. If we wished to use proper time we would set the
intrinsic coordinate $T$ equal to $q^{0}(t)/p^{0}$.
It is straightforward to show that in this case the resulting unique
trajectories in terms of this new intrinsic coordinate are
$$
Q^\mu(T) =
q^\mu - \frac{p^\mu}{ p^0} q^0 + p^\mu T.
$$
We will discuss on the physical significance of these results
in section 7.
%%%%%%%%%%%%%%%%%%%%%%%%%%%%%%%%%%%%%%%%%%%%%%%%%%%%%%%%%%%
\section{The Komar-Bergmann procedure as a gauge fixing.}
%%%%%%%%%%%%%%%%%%%%%%%%%%%%%%%%%%%%%%%%%%%%%%%%%%%%%%%%%%%%%%
Once two observers agree on the set of scalars $A^I$ to
use, we claim that they will describe the same physics if
their descriptions in their respective intrinsic coordinates coincide.

The results of the previous
section can be given a different perspective from the point of view of
gauge fixing. Indeed, given a metric $g$, the functions 
$G^{IJ}$ of section 5.2
are the solution of the four gauge fixing constraints 
\beq
\Phi^I := x^I -
A^I(W[g(x)]) = 0.
\label{gf}
\eeq

We employ here the usual definition of gauge fixing. Given a solution
of the Einstein equations in some given coordinate system we consider
all solutions obtainable from this solution through the action of all
finite gauge transformations generated by (\ref{thegen}) for arbitrary finite
$\xi$. These solutions lie on a gauge orbit.
Among these functionally different solutions we demand that
there exist only one for which (\ref{gf}) is identically satisfied.

Indeed, consider that $g$ is a metric  solution of the
constraints $\Phi^I = 0$.
Then an infinitesimally close metric in the same gauge orbit,
$$
g'
= g + {\cal L}_{\epsilon^\mu \partial_\mu}(g)
$$
can not be a
solution of the constraints. In fact, using (\ref{det}),
$$
A^I(W[g'(x)])
= A^I(W[g(x + \epsilon)]) = a^I_{g}(x + \epsilon) = a^I_{g}(x) +
\epsilon^\mu \partial_\mu a^I_{g}(x) \neq a^I_{g}(x) = x^I,
$$
that is,
$$
A^I(W[g'(x)]) \neq x^I
$$
for at least one value of $I$. Notice
that the solution of the constraints in each gauge orbit obviously
changes if we adopt a different set of scalars $\hat A^I$ as defined in
(\ref{hata}).

A complementary interpretation is available for fixing the
gauge. Since we have at our disposal finite gauge transformations
corresponding to finite changes of coordinates, we can find the in
general dynamical-solution-dependent gauge transformation which will
transform any given solution to one satisfying the gauge conditions. The
resulting gauge transformed solutions are by construction
invariants - i.e., observables.

In addition, the formalism provides
instructions on transforming from one set of observables  to another set. In the
context of Komar-Bergmann gauge fixing, these instructions amount to
implementing diffeomorphisms on the the Weyl scalar coordinates, as
expressed in (\ref{hata}).
Different choices of the functional form of the scalars $A^{I}$
will correspond to different sets of observables, and the formalism
tells us how to convert from one set to another. We will
exhibit this procedure in detail for the free relativistic particle.

Up to now we have verified that the constraints $\phi^I$
produce a good gauge fixing in the space of metric configurations, for
they select, at least locally, a single representative for each gauge
orbit. For the remainder of this subsection we will consider that the dynamics of the metrics is
given by GR and will study the role of the Komar-Bergmann constraints in
fixing the dynamics, that is, in the building of a solution of Einstein
equations starting from a well posed set of initial data. We will work
in a local system of coordinates ${x^\mu}$ such that $x^0$ has a
``time'' interpretation, that is, the surfaces defined by the constancy
of $x^0$ are spacelike. Also, we will formulate the dynamics in phase
space. It is worth remembering  that the scalars $A^I$ are
indeed functionals of a reduced set of the phase space variables, namely
$g_{ab}$ and $\pi^{ab}$, assuming, of course, that we are working
with solutions of the Einstein equations \cite{berg-kom60}.

The constraint structure of
canonical (ADM) 
GR is given by four primary constraints, which are the canonical
conjugate $P_\mu$ to the lapse $n^0 := n$ and shift $n^a$ variables, and
four secondary constraints, which are the so called Hamiltonians ${\cal
H}_\mu(g_{ab}, \! \pi^{ab})$ \cite{adm}. All constraints are first class and the
Poisson brackets between the Hamiltonians define a set of structure
functions $C_{\mu\nu}^\sigma$. The gauge generator in phase space
associated with infinitesimal diffeomorphisms was given in
\cite{pss:1997pr} (here we keep explicit the 3-space integration)
\begin{equation}
    G_{\xi}(x^0) = \int d^3\!x \left(P_\mu
\dot\xi^\mu + (
{\cal H}_\mu + n^\rho C^\nu_{\mu\rho} P_\nu) \xi^\mu \right),
\label{thegen2}
\end{equation}
for $\xi^\mu$ arbitrary descriptor
functions of the spacetime coordinates.
A good gauge fixing is one for which the vanishing at all times
of the equal time Poisson brackets
of the gauge fixing constraints with the gauge generator
eliminates all possible gauge transformation freedom, that is
$$
\{ \Phi^I, \, G_{\xi}(x^0) \} = 0 , \forall x^0 \
\Longrightarrow \ \xi^\mu = 0 \ ,
$$
This means that
\beq \det(\{A^I,
\, {\cal H}_\mu \}) \neq 0. \label{det2}
\eeq
The dynamical generator of time evolution, the Dirac Hamiltonian, is
\cite{wald84}
\begin{equation}
H_{\lambda} =
\int d^3\!x \ \big( n^\mu {\cal H}_\mu + \lambda^\mu P_\mu \big) =: H_c
+ \int d^3\!x \ \lambda^\mu P_\mu \ , \label{dirach}
\end{equation}
where $\lambda^\mu$ are Lagrange multipliers that must become determined
when the gauge fixing constraints are implemented. In fact, the time
stabilization of the gauge fixing constraints gives new constraints
\beq
{\dot \Phi^I} = \frac{\partial \Phi^I }{ \partial t} + \{ \Phi^I, \,
H_{\lambda} \} = \delta^I_0 - \int d^3\!x' \{ A^I, \, {\cal H}'_\mu \}
{n'}^\mu
= 0,
\label{gf2}
\eeq
(where primes indicate dependences with respect to the cordinates $x'$)
which, upon a new time stabilization,
\beq
{\ddot
\Phi^I} = - \Big\{ \{ A^I, \, H_c \}, \, H_c \Big\} - \int d^3\!x' \{ A^I, \,
{\cal H}'_\nu \} {\lambda'}^\nu = 0,
\label{detlambda}
\eeq
determines the
Lagrange multipliers, taking into account (\ref{det2}).

Notice that since by assumption the full set of constraints and gauge
conditions is second class, the Poisson bracket $\{ A^I, \, {\cal H}_\mu
\}$ possesses a matrix inverse, and (\ref{gf2}) may be solved for the
lapse and shift as functions of the non-gauge variables. It is
noteworthy that had the gauge fixing conditions not possessed an
explicit time dependence (in this case in $\Phi^0$), the lapse and
shift would have been zero! Our
constraints $\Phi^I$ thus conform with the general result
\cite{pss:1997pr,ps:95cqg}, already cited in section 4,
that the gauge fixing constraints for
generally covariant theories must always include an explicit time
dependence. In fact, the Komar-Bergmann constraints exhibit explicit
dependence, not only on the time coordinate, but on all spacetime
coordinates since each constraint solves for one coordinate.

We can obtain an equivalent explicit expression for the lapse and shift
taking into
account that the $A^{I}$ are scalars, and that we are working in a
coordinate system in which (we switch to greek indices for convenience)
$x^{\mu} = A^{\mu}$, so $ \partial_\nu A^\sigma =
\delta_\nu^\sigma $. Consequently
\beq
\delta A^\sigma =
\epsilon^\mu\partial_\mu A^\sigma = \xi^0 {\cal N}^\mu \partial_\mu A^\sigma +
\xi^a\partial_a A^\sigma = \xi^0 n^{\sigma} +\xi^{a} \delta^{\sigma}_{a}.
\eeq
On the other hand this infinitesimal transformation is generated by
(\ref{thegen}). Comparison gives the results
\beq
\{ A^\sigma(x^0, {\bf
x}), \, {\cal H}_\mu(x^0, {\bf x'}) \} = \bigl(\delta^0_\mu {\cal N}^\sigma (x^0, {\bf x'}) +
\delta^a_\mu \delta^\sigma_a \bigr) \delta({\bf x} - {\bf
x}').
\label{commut}
\eeq
These relations impose conditions on the functional forms of
$A^{\sigma}(W)$. Perhaps the most significant is that $A^{0}$ must be
chosen so that ${\cal N}^{0}$ is positive definite, but the requirement that
\beq
\{ A^{\mu},{\cal H}_{a} \} = \delta^{\mu}_{a}
\delta({\bf x} - {\bf
x}') . \label{Aa}
\eeq
is also non-trivial.
It follows from (\ref{commut}) that
\beq
{\cal N}^{\mu}(x^{0}, \vec x) = \{ A^{\mu}(x^{0}, \vec x),\int d^3\!x'
{\cal H}_{0}(x^{0}, \vec x') \}. \label{nmu}
\eeq
Care must be exercised in interpreting  (\ref{nmu}). It is actually a
constraint which expresses the canonical lapse and shift variables, see
(\ref{lapseandshift}), in
terms of the remaining variables. It is equivalent
to (\ref{gf2}),
whereas (\ref{Aa}) are just the spatial derivatives of the
Komar-Bergmann constraints.

Also, we can substitute (\ref{commut})  directly into the $\lambda$
dependent term in
(\ref{detlambda}) to determine the Lagrange multipliers. We
obtain
\beq
\lambda^0 = - n \Big\{ \{ A^0 , \, H_c \}, \, H_c \Big\} , \,
\lambda^a = - \Big\{ \{ n A^a + A^a , \, H_c \}, \, H_c \Big\} , \label{lamda0}
\eeq
and therefore the Dirac Hamiltonian can be written as
$$
H_D = \int d^3\!x
\left( n^\mu( {\cal H}_\mu - \Big\{ \{ A^0 , \, H_c \}, \, H_c \Big\}
P_\mu ) -
\Big\{ \{ A^a , \, H_c \}, \, H_c \Big\} P_a \right) \ .
$$

%%%%%%%%%%%%%%%%%%%%%%%%%%%%%%%%%%%%%%%%%%%%%%%%%%%%%%%%%%%%%%%%%%%%%%%
\subsection{Gauge fixing for
the free relativistic particle}

We conclude this
section with an illustration of gauge fixing and the associated
determination of lapse and Lagrange multiplier for the free relativistic
particle.
The preceeding discussion is applicable since the only
assumption used explicitly was that the coordinate time be set equal
to a scalar function of the non-gauge dynamical variables.

So let us investigate the implications of the gauge fixing
\beq
\Phi := t - f^{-1}(q^{0}) \approx 0,
\eeq
where $f^{-1}$ is a monotonically increasing but otherwise arbitrary
function. The function $f^{-1}$ plays the role of $A^{0}$. 
Now
according to (\ref{nmu}) the lapse must be given by
\beq
n^{-1}(t) = \{ f^{-1}(q^{0}(t)), \frac{1 }{ 2}(p^{2} +1)\} = p^{0} {d
f^{-1} \frac( q^{0}(t))}{ d q^{0}(t)} .
\eeq
But notice that since $f(f^{-1}(q^{0})) = q^{0}$, $1 = \dot f
\frac{d f^{-1} ( q^{0})}{ d q^{0}}$, and differentiating once more we find
that $0 = \frac{\ddot f}{ \dot f^{2}} + \dot f \frac{ d^{2} f^{-1}(q^{0})
}{ d(q^{0})^{2}}$.
Therefore
\beq
n(t)  = \frac{ \dot f (t)}{ p^{0}},
\eeq
which agrees with (\ref{dotN}). Finally, according to (\ref{lamda0})
\beq
\lambda = - n(t) \{ n(t) p^{0} \frac{d f^{-1}(q^{0}) }{ d q^{0}}, n(t) \frac{1}{
2}(p^{2} +1) \} = - n^{3}(t) p^{0} \frac{\frac{ d^{2} f^{-1}(q^{0})
}{ d(q^{0})^{2}}}{ \frac{d
f^{-1}(q^{0}) }{ d q^{0}}} = \frac{ \ddot f (t)}{ p^{0}}.
\eeq

%%%%%%%%%%%%%%%%%%%%%%%%%%%%%%%%%%%%%%%%%%%%%%%%%%%%%%%%%%%
\subsection{Degrees of freedom through the Komar-Bergmann method }
%%%%%%%%%%%%%%%%%%%%%%%%%%%%%%%%%%%%%%%%%%%%%%%%%%%%%%%%%%%%%%
We have
studied the role of the Komar-Bergmann gauge fixing constraints
(\ref{gf}) in two different frameworks. The first, in section 5.2,
was the space of spacetime metric histories with no
dynamical content; no dynamical stabilization algorithm was
invoked. Once
the scalars $A^I$ have been chosen, the gauge fixing (\ref{gf}) selects,
at least locally, a single metric in each gauge orbit. The global
question is left unanswered because we are not able to
rule out the
possible appearance of Gribov-type ambiguities.
The second framework, analyzed in this section, was the space of
solutions of Einstein equations.
We showed that the constraints (\ref{gf}) fix completely the Einstein
dynamics because the
stabilization algorithm fixed uniquely in (\ref{detlambda}) the Lagrangian
multipliers in the Dirac Hamiltonian (\ref{dirach}). Now let us
count the number of independent variables. The lapse and shift
variables are determined through (\ref{gf2}) in terms of the other
variables. Also, since the primary constraints $P_\mu$ are the
canonical conjugate variables to the lapse and shift variables, they
are determined as well --to be zero-- and so we have $2 \times 4
= 8$ variables already determined. At this point we are left with
$g_{ab}$ and $\pi^{ab}$ as independent variables, adding up to a total
of $12$. To these variables we must apply the four restrictions coming
from the secondary constraints ${\cal H}_\mu = 0$ (the Hamiltonian and
momentum constraints) and also the original Komar-Bergmann gauge
fixing constraints (\ref{gf}). We are left with $12-8 = 4$ independent
variables, corresponding to the two standard degrees of freedom of
General Relativity. Since the Einstein dynamics has been completely
fixed, that is, it has become a deterministic dynamics, we could
study the degrees of freedom as the freedom of setting the initial data
at, say, $x^0 = 0$. Giving values at that time to the four independent
variables will determine a solution with a unique physical content.
Giving other values to the four independent variables will determine a
physically distinct solution. So changing the initial values of the four
independent variables amounts to changing the gauge orbit; all the
metrics in the same gauge orbit define the same physics. The freedom
to give arbitrary values to the four independent variables is consistent
with the fact that, in the space of metrics, the Bergmann-Komar gauge
fixing constraints (\ref{gf}) select, at least locally, a single metric
in each gauge orbit.

\section{Observables}

We interpret observables in any generally covariant theory to be
those functions of dynamical variables which are invariant under
diffeomorphisms. In phase space formulations of generally
covariant theories this characterization must be altered to read
``invariance under diffeomorphism-induced transformations''. We
shall first present a concise general argument for a
Komar-Bergmann type construction. Then in the following
subsection we demonstrate exlicitly the invariant nature of
objects constructed using Komar-Bergmann gauge fixing and we
inquire into their physical observability. In the next subsection
we write down invariants for the free particle and we show explicitly
that they remain invariant under the action of the gauge group.
In this
context we also show that, contrary to initial expectations,
there is no  necessary  relation between invariants and
additional symmetries of the equations of motion, and we will
explain why.

\subsection{Komar-Bergmann type observables}

The primary ingredients in the Komar-Bergmann construction are an
intrinsic coordinate fixation using a scalar function of dynamical
variables, and a scalar function of variables expressed in these
coordinates. The idea that the specification of four independent
scalars could bring observables for GR is an old one. Besides the
work of Komar and Bergmann which is an elaboration of a
suggestion by G\'{e}h\'{e}niau and Debever that Weyl
scalars be used for this purpose \cite{gd56a,gd56b}, DeWitt \cite{dewitt62},
Isham and Kuchar
\cite{ishamkuchar}, Hartle \cite{hartle91}, and Marolf \cite{marolf95} have
also
advocated the use of scalars. Let us explore again, this
time in a formally precise way, why this procedure
delivers invariants.

We consider a generic generally covariant theory in which we have
dynamical variables, or functions of variables, which transform as
scalars under diffeomorphisms. Let $s(x)$ represent an independent
set of scalars equal in number to the dimension of spacetime. We
suppose they are of second differential order in the dynamical
fields, after imposition of the equations of motion, and can
therrefore be expressed in terms of phase space variables. We let
$Q(x)$ represent the full set of dynamical variables. The first
step in the construction is to set intrinsic coordinates $X$
equal to $s(x) =: a(x)$, where we suppose that $a$ is an
invertable coordinate transformation. We interpret $ X = s(x)$ as 
a dynamical
variable dependent coordinate transformation which depends only
implicitly on $x$ through the $x$ dependence of $s$. The geometric
variables obtained under this map are $\bar Q(X) = a^{*}
Q(x)$. But suppose we first undertake an arbitrary finite
coordinate transformation $x_{f} = f(x,Q)$, where we permit this
transformation to even depend on the dynamical variables. Under
this transformation $s_{f}(x_{f}) = s(x)$. Follow this
transformation with the transformation to intrinsic coordinates,
then because the $s$ are scalars:
\beq
X_{f} = s_{f} (x_{f})
= s(x)  =X. \label{barsf}
\eeq
This key result, $X_{f}(x)= X(x)$ shows that the numerical values of the intrinsic
coordinates are the same, and the resulting coordinate transformation
from $x$ to $X$ is identical, in spite of the indirect route. It
follows therefore that the map of geometric objects $a^{*}$ is
identical, and that the resulting geometrical objects expressed in
terms of intrinsic coordinates are identical, i.e., they are
invariant under the arbitrary coordinate map $f$.

The invariance we are discussing here is the usual notion of
invariance in any theory which possesses a local gauge symmetry. We
imagine we have a solution of the equations of motion which we have
expressed in an arbitrary coordinate system. Thus we want to consider
the objects we construct by going to intrinsic coordinates as phase
space functions of these original variables. They undergo non-trivial
variations engendered by our symmetry generator $G[\xi]$. On the
other hand, if we were to take a solution which already satisfies the
gauge fixing condition (rather than our invariant function of solutions)
 and perform a gauge transformation on it, we
would of course obtain a solution which no longer satisfies the gauge
condition! We will illustrate these ideas in detail in the next
section using the
relativistic free particle.

Let us now consider the physical measurement of the full
four-dimensional metric tensor in an intrinsic coordinate system.
There is in principal a well-defined procedure at our disposal. It
relies on a device first envisioned by Peter Szekeres which he has
called a ``gravitational compass''\cite{sz65}. It consists of a
tetrahedral arrangement of springs. By measuring the stresses in
the springs one can determine components of the curvature tensor.
In the vacuum case three compasses will suffice to determine all
of the local components of the Weyl tensor. Four compasses are
required to determine the full local Riemann curvature tensor in
the presence of matter sources. These measurements can in
principal be used to establish the intrinsic coordinate system
fixed by the Weyl scalars. Supplemental measurements of distances
using light ranging will then determine components of the metric
in this coordinate system.
%%%%%%%%%%%%%%%%%%%%%%%%%%%%%%%%%%%%%%%%%%%%%%%%%%%%%%%%%%%%%%%%
%%%%%%%%%%%%%%%%%%%%%%%%%%%%%%%%%%%%%%%%%%%%%%%%%%%%%%%%%%%%%%%%
\subsection{Observables for the free relativistic particle}
%%%%%%%%%%%%%%%%%%%%%%%%%%%%%%%%%%%%%%%%%%%%%%%%%%%%%%%%%%%%%%%%
%%%%%%%%%%%%%%%%%%%%%%%%%%%%%%%%%%%%%%%%%%%%%%%%%%%%%%%%%%%%%%%

We now consider the construction of invariants for the free
relativistic particle in the manner just described. Actually the
job was already completed in section 5.3. The idea is that we
choose a scalar function of the dynamical variables, and then use
this scalar to define a parameter transformation $T(t)$.
Then we can construct invariants out of all components of the
spacetime position and the lapse by setting $Q^{\mu}(T) =
q^{\mu}(t(T))$ and $N (T) = N(t(T)) \frac{dt}{
dT}$. We will explicitly construct classes of
gauge invariants for the free particle corresponding to a wide
class of gauge choices.

We first consider invariants using the intrinsic coordinate 
\beq
T= f^{-1}(q^0(t)), \label{fgauge} 
\eeq 
where $f^{-1}$ is a
monotonically increasing but otherwise arbitrary function. We
found in obtaining (\ref{qumut}) and (\ref{Nt})  that we 
did not need to solve
explicitly for $t$ in terms of $T$. We merely substituted
the $t$ dependent term in the general solution for $q^{0}(t)$
into the expression for $q^{a}(t)$, obtaining  
\beq 
Q^a(T) = q^a(t(T)) = q^a_0 + \frac{p^a }{ p^0} (f(T)
-q^0_0). \label{parinv} 
\eeq 
Similarly we solved for $\frac{dt}{ d T}$ to find
\beq
 N (T) = \frac{\frac{df(T)}{ dT} }{p^{0}}. \label{Ninv}
\eeq 
These are our putative invariants. 

It will be useful to rewrite these invariants in terms of the solution trajectories $n(t)$ and $q^\mu (t)$ given by (\ref{Nt}) and (\ref{qt}).  Substituting for the initial values $q^\mu_0$ we find
\beq
Q^a(t) = q^a (t) + \frac{p^a}{p^0} \left(f(t) - q^0 (t) \right), \label{barql}
\eeq
while $N (t) = \frac{1}{p^0} \frac{df(t)}{dt}$ is unchanged.

Now let us examine the variations of these objects under an
arbitrary infinitesimal canonical gauge transformation. We shall demonstrate invariance in two equivalent ways. In the first procedure we express all phase space variables in terms of initial values, and Poisson brackets will be computed in terms of these initial value phase space coordinates. We note
that the only relevant nonvanishing variations engendered by
\beq
G_{\xi}(t) = \frac{1 }{ 2} \xi(t) (
p^2 + 1) + \dot \xi(t) \pi(t),
\eeq
are (since none of the
invariants depend on $n$)
\beq \delta q^\mu = (\xi(t) - t \dot
\xi(t))p^\mu.
\eeq
Therefore $Q^0(t)$  and $ N ( t)$ are  trivially invariant,
while $Q^a(t)$ is invariant since the $q^\mu_0$ coordinates appear in
the combination $q^a_0 - \frac{p^a}{ p^0} q^0_0$.

A second equivalent procedure available to us is to compute Poisson brackets at the time $t$ with respect to the canonically evolved phase space variables at time $t$. Thus the relevant non-vanishing variations generated at time $t$ are
\beq
\delta q^a (t) = \Big\{ q^a (t), \frac{1}{2} \xi(t)   p^2 (t)  + \dot \xi (t) \pi (t) \Big\}_{y (t)} = \dot \xi (t) p^a (t) = \dot \xi (t) p^a.
\eeq
Therefore, referring to (\ref{barql}),
\beq
\delta Q^a (t) = \delta q^a (t) - \frac{p^a}{p^0} \delta q^0 (t) = \xi (t) ( p^a - p^a ) = 0.
\eeq

Notice that these invariants are in general dependent on $t$, and
{\it not} constants of the motion. The independence of gauge and
time evolution is made strikingly evident in this example. Notice
also that our observables are also invariant under the action of
the gauge fixed Hamiltonian (\ref{fham}). At first sight this may
appear to be a contradiction since we have simply expressed an
arbitrary solution of the equations of motion in terms of
intrinsic coordinates. We might well ask: should this solution
satisfy the equations of motion? If it did satisfy the equations
of motion then it's Poisson bracket with the Hamiltonian would
{\it not} vanish. But the apparent contradiction is resolved when
one realizes that $Q^a( t)$ given by (\ref{barql}) exhibits explicit time
dependence.
Thus we have written the
invariants as the sum of a part constructed with solution
trajectories plus a part which contains an explicit $t$
dependence. Therefore 
\beq 
\frac{dQ^{\mu}(t)}{ dt} =
\frac{\partial Q^{\mu}(t) }{ \partial t} + \{ 
Q^{\mu}(t), H_{D} \} = \frac{p^{\mu} }{ p^{0}}\dot f + 0 =
\frac{p^{a}}{ p^{0}}\dot f,
\eeq
which agrees with (\ref{qmudot}).

There is a widespread mistaken notion in
the literature that gauge invariants in generally  covariant
theories must be constants of the motion. (See the Conclusion for further discussion and references). Our gauge invariants
for the free particle are a counterexample. And since they are
not constants of motion, they should not be expected to generate
symmetries of the equations of motion.

There do exist invariants for the free particle which {\it are} constants of the
motion, and it will be instructive to examine some of them. One
such class can be obtained even before adopting intrinsic time.
Consider the solutions
(\ref{thetraj}). They satisfy \beq
 q^{\mu}(t) - \frac{p^{\mu} }{ p^{0}}  q^0(t) =
 q^{\mu}(0) - \frac{p^{\mu}}{ p^{0}}  q^0(0)
\eeq that is, \beq q^{\mu}(t) - \frac{p^{\mu}}{ p^{0}}  q^0(t)
\label{boosts} \eeq
 are constants of motion (the time component vanishing)
with no explicit time dependence. One can check that they are also
gauge invariant quantities.

Notice that these very same gauge invariant quantities
(\ref{boosts}) can be presented, when described with the
intrinsic coordinates, with explicit time dependence. Indeed,
this can be achieved by isolating the new initial conditions on
the trajectory $(\ref{parinv})$, 
\beq 
Q^\mu(0) = q^{\mu}_0 +
\frac{p^{\mu}}{ p^{0}} (f(0) - q^{0}_0), 
\eeq 
which are evidently
gauge invariant quantities. Then the trajectory can be expressed
as 
\beq
 Q^\mu( t) = Q^{\mu}(0) + \frac{p^{\mu} }{ p^{0}} (f(t) -
f(0)), 
\eeq 
which identifies the time-dependent constants of the
motion
\beq L^{a} :=  Q^{a}(t) - \frac{p^{a}}{ p^{0}}
f(t). 
\eeq 
These constants of
the motion are just the constant initial values,
\beq L^{a} = q^a - \frac{p^a }{ p^0} q^0. \eeq

Constants of the motion are generators of symmetries of the
equations of motion, and map solutions into solutions. It is not
unusual for a constant of motion to be time dependent, as are for
example the Noether constants of motion associated with Galilean
boosts. In fact, the $L^{a}$ are generators of Lorentz boosts as
can be straightforwardly shown. We notice that nothing analogous
to these boosts generators in a gauge fixed theory can exists in
vacuum general relativity because there exists no dynamical
symmetries beyond general covariance.

We have pursued this example in some detail to make a significant
point. C. G. Torre has asserted that in general relativity there
can exist no observables built as spatial integrals of local
functions\cite{torre93}. In fact the Komar-Bergmann construction
in the case of the free particle provides local observables. It
is true that observables commute with the Hamiltonian
constraint.   But whereas constants of the motion generate
symmetries and map solutions onto solutions, non-constant
invariants do {\it not} map solutions onto solutions. What Torre has actually proven, and it is in our view no less significant, is that in vacuum general relativity there exists no {\it constant} } 
in time observables built as spatial integrals of local phase space functions!
The Komar-Bergmann observables are indeed local in
both space and time. This follows from the fact that the
intrinsic coordinates are local functions of the
spatial metric and conjugate momenta and spatial derivatives
thereof. These are in turn algebraic functions of spatial and
time coordinates. Thus the mapping from arbitrary spacetime
coordinates to intrinsic coordinates is local, as is the
inversion map. In addition, the metric components in the original coordinate
patch are local functions of the coordinates, and they therefore
remain local functions when expressed in terms of the intrinsic
coordinates.

It is also clear from this free particle example that given {\it
any} parametrization of the particle world line there is a
corresponding set of invariants, corresponding to the choice of
the function $f$! Are these invariants measureable, and
therefore observable? Indeed they are in the context of flat
space time where we assume we have coordinate clocks distributed
throughout space. These coordinate clocks are usually set to run
with the gauge fixing condition $f(t) =  t$, so $q^{0} = t $. The
reading of the clock constitutes a partial observable in the
sense of Rovelli\cite{rovelli01a}. Complete observables are
correlations between partial observables, and the correlations
are fixed by the theory. The observables cited above admirably
fit this description when we take into account that choices of
the gauge fixing function $f$ merely correspond to differing
instructions on adjusting the rate of rotation of the clock hands
with respect to the flow of Minkowski time.

\subsection{Invariants, gauge fixing and Dirac brackets}

Since we are now able to implement finite diffeomorphism-induced
gauge transformations we have at our disposal a standard procedure
for producing gauge invariants through the imposition of gauge
conditions. After describing the general method
we will apply it to the relativisitic particle, and then comment on
the general relationship between invariants,
gauge fixing, and Dirac brackets.

As before we let $y(t)$ represent the set of canonical
solution trajectories corresponding to the Dirac Hamiltonian
$H_{\lambda}$. But we shall alter our previous notation somewhat and
represent a finite gauge transformed trajectory with descriptor $\xi$
and group parameter $s = 1$ as $y_{\xi}(t)$. Let us impose
gauge conditions $\chi(y(t)) = 0$. We achieve this condition by
performing the appropriate $y(t)$ dependent gauge
transformation on $y(t)$, described by a $y(t)$
dependent descriptor $\xi$. Thus the objects $y_{
\xi(y(t))}(t)$ are manifest gauge invariants.

In the case of the free relativistic particle the descriptor is fixed
by the gauge condition
\beq
f(t) = q^{0}_{ \xi}(t) = q^{0}(t) + p^{0} \xi(t),
\eeq
resulting in $\xi(y(t),t) = \frac{1}{ p^{0}} (f(t) - q^{0}(t))$.
Thus we recover the gauge invariants displayed above:
\beq
q^{a}_{ \xi(q(t),p(t),t)} = q^{a}(t)
+\frac{p^{a} }{ p^{0}} \left( f(t) - q^{0}(t) \right) =
q^a + \frac{p^a}{ p^0} (f(t) -q^0).
\eeq

We complete this section noting that Dirac brackets are simply the
ordinary Poisson brackets of our invariants. For example, employing
the inverse matrix $M^{mn}$ in (\ref{Mmn}) we find
\bea
\{N(t),q^{a}(t) \}^{*} &=& - \{ N(t),\psi_{2}(t)\} M^{21}(t) 
\{\psi_{1}(t), q^{a}(t)\} \nonumber \\
&=& - \frac{N(t)}{(p^{0})^{2}}p^{a} = \{ N(t), Q^{a}(t) \},
\eea
where in the last equality (\ref{parinv}) and (\ref{Ninv}) have been used. 

%%%%%%%%%%%%%%%%%%%%%%%%%%%%%%%%%%%%%%%%%%%%%%%%%%%%%%%%%%%%%%%%
%%%%%%%%%%%%%%%%%%%%%%%%%%%%%%%%%%%%%%%%%%%%%%%%%%%%%%%%%%%%%%%%

\section{Conclusions}

Our focus throughout this paper has been the distinction between time
evolution and diffeomorphism gauge symmetries in generally covariant
theories, and the startling physical consequences of this
distinction. Time evolution is of course the mapping of initial data
to produce solution trajectories. Diffeomorphism gauge transformations
map entire solution trajectories into solution trajectories. The
distinction is obvious in the usual configuration-velocity space
formulation. Clearly if, for example, $g_{\mu \nu}(x^{0}, \vec x)$
represents a solution of Einstein's equations, then under an
infinitesimal coordinate transformation ${x'}^{\mu} = x^{\mu} -
\epsilon^{\mu}(x^{0}, \vec x)$, the corresponding active variation of
$g_{\mu \nu}(x^{0}, \vec x)$ is just the Lie derivative
\beq
{\cal L}_{\epsilon} g_{\mu \nu} = g_{\mu \nu, \alpha}
\epsilon^{\alpha} + \epsilon^{\alpha}_{,\mu} g_{\alpha \nu} +
\epsilon^{\alpha}_{,\nu} g_{\mu \alpha}.
\eeq
Clearly there is a different variation at each time $x^{0}$. It is in
the transition to phase space that one can easily lose one's way.

Fortunately we now have at our disposal a concrete realization of the
full diffeomorphism-induced gauge transformations group in phase space.
And the distinction between time evolution and gauge is made even more
transparent when lapse and shift functions are retained as canonical phase space
variables. There is an essential distinction between the Hamiltonian
and the generator of gauge transformations. They are similar in
appearance, but in the Hamiltonian we have the arbitrary coordinate
functions $\lambda^{A}$, whereas in the gauge generator these
coordinate functions are replaced by the canonical variables
$\dot n^{A}$. 

The misidentification of evolution and gauge has led to the often
repeated assertion that gauge invariants in generally covariant
theories must be constants of the motion. It is true that
the Poisson brackets of invariants with all of the first class
constraints in a generally covariant theory must vanish (since these
constraints all appear multipying arbitrary functions in the
generator $G(t)$). But this assertion fails to take into account any
explicit time dependence  (as opposed to implicit time dependence,
that is, time dependence appearing in
canonical variables) in the invariants of the theory. We have shown
that such an explicit time dependence arises in any acceptable gauge
fixing. It is compulsory in order to uniquely fix a solution on the
gauge orbit.
We have shown in detail how the Hamiltonian dynamics in generally
covariant theories accomodates time-dependent gauge fixing. In
particular, the dynamics is not frozen; time evolution is non-trivial
after the imposition of time-dependent gauge conditions.

We have explored a special class of gauge conditions in which scalar
functions of dynamical variables are selected to define intrinsic
coordinates. Komar and Bergmann made the original concrete proposal of
this type in general relativity with the suggestion that Weyl
curvature scalars serve as intrinsic coordinates in vacuum general
relativity. (We have shown that Weyl scalars may be used for this
purpose also when material sources are present.) We demonstrated why
this procedure produces invariants.

Throughout this paper we have illustrated our discussion with
applications to the free relativistic particle. This is a dynamical
model for which we possess the general solution but which
nevertheless possesses a highly non-trivial symmetry structure in
phase space. We imposed an analogue of the Komar-Bergmann gauge
fixing, recovered a non-trivial dynamics due to the explicit time
dependence of this gauge choice, and ultimately displayed
diffeomorphism invariants which were not constants of the motion. We
also displayed constant invariants which were found to generate
Lorentz boosts of the particle world lines. We pointed out that a
theorem due to Torre can be reinterpreted as a demonstration that in
vacuum general relativity, as opposed to the relativistic free
particle, there can exist no constant (of the
motion) diffeomorphism
invariants constructed with spatial integrals of local canonical variables since
in vacuum general relativity there exists no dynamical symmetries
beyond general covariance.

We interpret invariants as observables. That the invariants displayed
for the free particle are indeed observable we think is beyond dispute
since they represent correlations observed in daily practice in modern
laboratories. In the language proposed by Rovelli, the time $t$ for
the free particle is a partial observable, recorded by a suitably
regulated clock, and our observables are complete in the sense that
for a choice of the regulating function $f$ the theory gives a
univocal prediction for particle spacetime position.

The invariants produced in the Komar-Bergmann gauge fixing in
general relativity enjoy a
similar status. Intrinsic spacetime coordinates are in principle
measureable and constitute partial observables. Given these
coordinates for a specific spacetime the theory univocally predicts
spacetime distances along arbitrary routes. And schemes exist, using
light ranging, for example, for measuring these spacetime distances. The
invariants resulting from Komar-Bergmann gauge are a univocal
correlation for these measurements, and therefore constitute complete
observables.

The assertion that diffeomorphism invariants must be constant in
time has a long and distinguished history, and is traceable at least
as far as Komar and Bergmann \cite{kb62}. Yet these authors explicitly
note situations, namely in regard to the use of intrinsic coordinates,
in which invariants display time dependence \cite{berg61b}. Rovelli
has explicitly addressed comparable apparent contradictions when on
the one hand he states that ``physical observables must be invariant
under evolution in $t$'' but points out that such a statement is
``ill posed, because it confuses evolution with respect to coordinate
time $t$ and physical evolution'' \cite{rovelli01a}.  In all of the works
cited the
paradox is resolved through the method of coincidences, or
eqivalently, intrinsic coordinates. The view
apparently espoused by these authors is that we can and should
distinguish between intrinsic time, on which variables might depend, 
and our initially arbitrarily chosen coordinate time. Invariants must
be independent of this latter choice. We fully agree. We have given a formal
elaboration of this distinction in our enlarged phase space in which
the full four-dimensional diffeomorphism-induced symmetry group is
realized as a canonical transformation group.

We close with some comments about the implications of this work for
an eventual quantum theory of gravity. The implications are profound.
A non-quantum evolutionary parameter which we should interpret as the
time will appear naturally in a Heisenberg picture formalism in which
states are functionals of the three-metric - and perhaps also of lapse
and shift. The time dependent invariants which appear in the
Komar-Bergmann gauge fixing may be promoted to operators (recognizing
as always that factor ordering amibiguities may arise). These
operators represent the full four dimensional metric in intrinsic
coordinates, and the full metric will therefore be subject to time
dependent fluctuations. Invariants have recently been constructed for a class of classical Bianchi type I cosmological models \cite{hss04}, and work is underway investigated the significance of these invariants in quantum cosmology.
%%%%%%%%%%%%%%%%%%%%%%%%%%%%%%%%%%%%%%%%%%%%%%%%%%%%%%%%%%%%%%%%%%%%%
\section{Acknowledgements}
We thank the Department of Physics at the University of Florence, and
particularly Giorgio Longhi for
hospitality during the first stages of this work. We also thank
Luca Lusanna for many stimulating discussions on the subject. DCS
thanks the Department of Physics at Syracuse University for its
hospitality. He thanks Josh Goldberg for his critical reading of an
earlier draft of this work, and many incisive comments. He also
thanks A. P. Balachandran and Don Marolf for stimulating and
enlightening discussions, as well as James Friedrichson, Josh Helpert,  Larry Shepley, and Allison Schmitz for spirited debate. DCS acknowledges recent financial support from the Priddy Foundation. 
JMP acknowledges financial
support by CICYT, AEN98-0431, and CIRIT, GC 1998SGR.
%%%%%%%%%%%%%%%%%%%%%%%%%%%%%%%%%%%%%%%%%%%%%%%%%%%%%%%%%%%%%%%%%%%%%
%%%%%%%%%%%%%%%%%%%%%%%%%%%%%%%%%%%%%%%%%%%%%%%%%%%%%%%%%%%%%%%%%%%%%
%%%%%%%%%%%%%%%%%%%%%%%%%%%%%%%%%%%%%%%%%%%%%%%%%%%%%%%%%%%%%%%%%%%%%

%%%%%%%%%%%%%%%%%%%%%%%%%%%%%%%%%%%%%%%%%%%%%%%%%%%%%%%%%%%%%%%%

%%%%%%%%%%%%%%%%%%%%%%%%%%%%%%%%%%%%%%%%%%%%%%%%%%%%%%%%%%%%%%%%
%%%%%%%%%%%%%%%%%%%%%%%%%%%%%%%%%%%%%%%%%%%%%%%%%%%%%%%%%%%%%%%%
\end{document}